# Influence of host matrices on krypton electron binding energies and KLL Auger transition energies

A.Kh. Inoyatov[a,b], L.L. Perevoshchikov [a], A. Kovalík[a,c], D.V. Filosofov[a], Yu.V. Yushkevich[a], M. Ryšavý[c], B.Q. Lee[d], T. Kibédi[d], A.E. Stuchbery[d], V.S Zhdanov[e],

[a] *Laboratory of Nuclear Problems, JINR, Dubna, Moscow Region, Russian Federation*
[b] *Institute of Applied Physics, National University, Tashkent, Republic of Uzbekistan*
[c] *Nuclear Physics Institute of the ASCR, CZ-25068 Řež near Prague, Czech Republic*
[d] *Department of Nuclear Physics, RSPE, The Australian National University, Canberra, ACT 0200, Australia*
[e] *Nuclear Physics Institute, Almaty, Kazakhstan*

**Abstract**

The low-energy electron spectra emitted in the radioactive decay of the $^{83}$Rb and $^{83}$Sr isotopes were measured with a combined electrostatic electron spectrometer. Radioactive sources used were prepared by ion implantation of $^{83}$Sr into a high purity polycrystalline platinum foil at 30 keV and by vacuum-evaporation deposition of $^{83}$Rb on the same type of foil. From the measured conversion electron spectra, the electron binding energies (referenced to the Fermi level) for the K, $L_1$, $L_2$, $L_3$, $M_1$, $M_2$, and $M_3$ shell/subshells of krypton in the platinum host were determined to be 14 316.4(12), 1 914.3(9), 1 720.3(9), 1 667.6(9), 281.5(9), 209.6(13), and 201.2(15) eV, respectively, and those for the evaporated layer were observed to be lower by 0.7(1) eV. For both host matrices, values of 2.3(2), 4.6(2), 1.7(2), 1.3(2), and 3.2(3) eV were obtained for the krypton K, $L_1$, $L_2$, $L_3$, and $M_1$ natural atomic level widths, respectively. The absolute energies of 10 838.5(9) and 10 839.5(10) eV were measured for the $KL_2L_3(^1D_2)$ Auger transition in krypton implanted in Pt and generated in the evaporated rubidium layer, respectively. A value of 601.0(8) eV was measured for the energy difference of the $KL_2L_3(^1D_2)$ transitions in Rb and Kr in the Pt host. Multiconfiguration Dirac-Fock calculations of the krypton KLL transition energies and intensities were also performed.

## 1. Introduction

Experimental investigations of the influence of the atomic environment on absolute binding energies of core electrons in medium heavy elements by means of "classical" electron spectroscopy methods (e.g. such as X-ray photoelectron spectroscopy (XPS)) are very rare also due to their technical limitations. This is especially true for the study of solid-host effects on absolute core level binding energies in noble gases, which are chemically inert and are the only free atoms. Thus, e.g., in the pioneer works [1,2] the binding energies of core electrons in Ne, Ar, Kr, and Xe implanted in Cu, Ag, and Au were investigated by X-ray photoemission and were found to be 2-4 eV smaller in magnitude than the corresponding binding energies obtained from gas-phase measurements. The aim of these measurements was to study contributions from both the change in the self-consistent potential experienced by the core levels upon implantation (the initial state) and the polarization of metal-host electrons upon photoionization (the final state) which are thought to be responsible for the observed shifts. The most studied core levels for which experimental data are published in Ref. [1,2] were K (Ne), $L_3$ (Ar), $M_3$ (Kr), and $M_5$ (Xe), i.e. atomic levels with relatively low binding energies (several hundred eV) easily accessible by standard XPS-apparatus.

The "low energy nuclear electron spectroscopy" used, e.g., in the present work exhibits considerable advantages in this field such as: i) a high signal-to-background ratio; ii) a wider spectrometer energy range (from "0" to 50 000 eV in the case of our combined electrostatic electron spectrometer [3]) enabling one to study much deeper core levels; iii) various methods for incorporation of parent isotopes of daughter noble gases into different solid hosts; iv) a possibility to study trace amounts of elements; v) electrons are emitted only by radioactive atoms, etc.

It should be noted that experimental information on the influence of the atomic environment on absolute core level binding energies in noble gases plays an essential role, for example, in the preparation of stable electron energy calibration standards for some significant physical experiments also in the field of high energy physics and in electron spectroscopy in general.

There is also a lack of experimental data on the influence of atomic environments on the KLL Auger spectra especially for medium and heavy elements. This information is important for both basic research in this field and interpretation of weak effects in complex experimental Auger electron spectra.

In the present paper we report results of our experimental investigation of environmental effects on the electron binding energies, the absolute KLL Auger transition energies, and natural atomic level widths in krypton generated in two different solid hosts. For this study, radioactive sources of both $^{83}$Sr ($T_{1/2}$=32.4 h, Fig. 1) prepared by ion implantation at 30 keV into a high purity polycrystalline platinum foil and $^{83}$Rb ($T_{1/2}$=86.2 d, Fig. 1) produced by vacuum evaporation on the same type of Pt foil were used.

Our investigations were performed in the frame of the development of a new technique for the preparation of super stable calibration $^{83}$Rb/$^{83m}$Kr electron sources [5,6] for the KATRIN neutrino mass experiment [7].

## 2. Experimental

### 2.1. Source preparation

Radioactive isotopes of strontium and rubidium were obtained by spallation of metallic yttrium by 300 MeV protons from the internal beam of the phasotron particle accelerator at the JINR, Dubna, Russia. After "cooling" for three days, the irradiated target (1 g weight) was dissolved in the concentrated nitric acid. Strontium isotopes were chemically separated from the target material and other elements using "Sr resin" (TrisKem International). An additional purification of strontium was then carried out on a cation-exchange chromatography column (70 mm length, 2 mm diameter, A6 resin) also in a nitric acid medium. Afterwards, the strontium fraction ($^{82}$Sr ($T_{1/2}$=25.3 d), $^{83}$Sr ($T_{1/2}$=32.4 h), $^{85}$Sr ($T_{1/2}$=64.9 d)) obtained was used for the electron source preparation.

Chemical separation of the rubidium fraction ($^{83}$Rb ($T_{1/2}$=85.2 d), $^{84}$Rb ($T_{1/2}$=32.8 d)) from the target material and other elements was also performed on the above "Sr resin" followed by twofold purification on the cation-exchange chromatography column in a nitric acid medium.

The mass separation of the strontium isotopes was performed on a mass separator at the JINR, Dubna. At the same time, the strontium ions were embedded at the energy of 30 keV into the high purity polycrystalline platinum foil. Prior to use, the surface of the Pt foils was cleaned only by alcohol. A part of the foil containing the Sr isotope with the atomic mass number A = 83 was cut out and used for the electron spectrum measurements. The size of the "active" spot was about 2x2 cm$^2$. The activity of the $^{83}$Sr isotope in the source upon the preparation was 11.7 MBq.

According to the simulations [5] performed for the implantation of $^{83}$Rb ions at 30 keV into the same type of platinum foil as used in our experiment, the mean projected range and standard deviation of the distribution of the implanted atoms along the depth of the foil amounted to about 9 and 4 nm, respectively. Real circumstances of the implantations [5] such as the zero ion incident angle (relative to the source foil normal), polycrystalline structure of the Pt foil, and an adsorbed surface contamination layer of water and hydrocarbons (so called "rest gas layer") were represented in the simulations by an additional 3 nm thick pure carbon layer on the foil surface. These experimental conditions were very close to our ones. It was, moreover, found in Ref. [5] that a portion of 5-10% of the incident $^{83}$Rb ions gets stopped in the "rest gas layer", which represents a non-metallic environment. As surface of the Pt foil was not cleaned (by ion sputtering or any other means) after the $^{83}$Rb ion collections, a certain portion of $^{83}$Rb thus remains in this non-metallic environment. This feature also pertains to our $^{83}$Sr source. After the collection of the $^{83}$Sr ions, the source was exposed to air during its transfer to the electron spectrometer. The $^{83}$Sr source surface was also not cleaned before measurements of electron spectra. Thus a portion of $^{83}$Sr ions stopped on the Pt foil surface in the non-metallic environment were bound with atoms of oxygen in all possible forms (oxides, hydroxides, carbonates, hydrocarbonates, etc.) and had the oxidation number +1 (see below).

   Thermal evaporation deposition of rubidium on the Pt foil took place at 900 °C. Prior to the use, the surface of the foil was cleaned by alcohol. An aqueous solution of a carrier free rubidium nitrate was transferred to an annealed Ta evaporation boat and dried up. To remove possible volatile organic compounds, the Ta evaporation boat with deposited activity was first heated at lower temperature for about 30 s. The source backing was shielded throughout this procedure. During the source evaporation, the source backing rotated around its axis at a speed of 3000 turns/min at a distance of 10 mm from the Ta evaporation boat to improve homogeneity of the evaporated layer. Activities of the $^{83}$Rb and $^{84}$Rb isotopes in the prepared source of 8 mm diameter were 2.2 and 0.3 MBq, respectively.
   The exact "chemical state" of the deposited trace amounts of $^{83}$Rb on the surface of the Pt source backing in vacuum was unknown. However, the source was transferred to the electron spectrometer in air after the preparation and thus the evaporated layer was exposed to it. Due to extreme rubidium reactivity to air, ions of $^{83}$Rb were bound with atoms of oxygen in all possible forms (oxides, hydroxides, carbonates, hydrocarbonates, etc. of different proportions) and had the oxidation number +1. This statement is based on: (i) specific chemical properties of rubidium, (ii) its known macro-chemistry, (iii) the inner self-consistency of the physicochemical methods used for the preparation of the source, and (iv) conditions of its treatment. After the EC decay of $^{83}$Rb, the daughter $^{83}$Kr atoms were stabilized in the above rubidium matrix and apparently they were not charged.

### 2.2. Measurements and energy calibration

   Electron spectra were measured in sweeps using the above mentioned combined electrostatic electron spectrometer [3] consisting of a retarding sphere followed by a double-pass cylindrical mirror energy analyzer. The choice of the absolute spectrometer energy resolution and the spectrum scanning step depended on the intensity of the radioactive source being measured and the complexity of the inspected spectrum. Examples of the measured spectra are shown in Figs. 2-5.
   Altogether 14 low energy conversion electron lines (listed in parentheses) of nuclear transitions in $^{169}$Tm with energies $E_\gamma$ = 8.41008(21) [9] ($M_{1,2}$, $N_{1,3}$), 20.74378(10) [10] ($L_{1-3}$, $M_{1-3}$, $N_{1-3}$), and 63.12081(5) keV [11] (K) along with five lines (K, $L_{1-3}$, $M_1$) of the 14.41300(15) keV [11] nuclear transition in $^{57}$Fe were used for calibration of the spectrometer energy scale. Energies of the calibration lines related to the Fermi level $E_F(i)$ ($i$ is the atomic subshell index) were evaluated as $E_F(i) = E_\gamma - E_{b,F}(i)$ making use of the experimental Fe and Tm electron binding energies $E_{b,F}(i)$ [12] related to the Fermi level. Experimental uncertainties from 0.4 to 1.6 eV and from 0.4 to 0.9 eV are quoted in Ref. [12] for thulium and iron electron binding energies, respectively. Thus in all cases the $E_{b,F}(i)$ uncertainties fully dominate uncertainties of the above nuclear transition energies. Therefore uncertainties in the calibration line energies were almost identical with those of the relevant electron binding energies.

In the case of Fe, the electron binding energies [12] are given for an oxide but the chemical form for Tm is not specified. Chemical techniques applied in the preparation of our calibration sources by vacuum evaporation on polycrystalline carbon substrate should guarantee the oxide state for both elements. For the case that it is not true, we took into account maximum measured chemical shifts of the outer subshell electron binding energies of about 2 and 4 eV [13] among different chemical states of Tm and Fe, respectively. We found that the influence of these shifts on the energy calibration is well below the standard deviations of the measured absolute energies of the studied electron lines (quoted in Tables 1, 7 and in the text).

*2.3. Spectra evaluation*

The measured electron spectra were evaluated using the approach and the computer code described in Ref. [14]. The individual spectrum line shape was expressed by a convolution of a Gaussian ("spectrometer response function to monoenergetic electrons") and an "artificial" function which describes the natural energy distribution of the investigated electrons leaving atoms (Lorentzian) distorted by manifold inelastic scattering of the electrons in the source material (surface and volume plasmon excitations, shake-up/-off, atomic and lattice excitations, etc.). In order to find the proper form of the complex discrete energy loss peak and its long low energy tail (going down to the "zero" energy), multiple fitting of the measured electron spectrum with random variations (Monte Carlo method) of the shape of the fitted lines in this "energy loss region" within the pre-set shape limits was applied. It should be noted that the position and width of the discrete energy loss peak (DEL – see, e.g., Figs. 3-5) depend on the source material, while its intensity relies on a ratio of the energy-dependent mean free path for inelastic electron scattering and the effective source thickness. (Contrary to the inelastically scattered electrons, the electrons which left the electron source without any energy loss create a so-called zero loss (or no-loss) peak. This peak can be described by a simple convolution of a Gaussian and a Lorentzian resulting in a Voigt function.)

In the evaluation, the fitted parameters were the position, the height, and the width (of the no-loss peak) of each spectrum line, the constant background and the width of the spectrometer response function. Results of the evaluations are shown in Figs. 3 and 4 (continuous lines), in Tables 1,6,7 and in the text. The quoted uncertainties are our estimates of standard deviations ($\sigma$).

**3. Results and discussion**

As can be seen from Figs. 3-5, the discrete energy loss peaks (DEL) of the conversion electron lines for the $^{83}$Rb source prepared by vacuum evaporation on the Pt foil are higher than those for the $^{83}$Sr source prepared by ion implantation into the same foil. On the other hand, the low energy tails of the spectrum lines converge faster to the background level for the evaporated source. This feature is caused by manifold scattering of the electrons in the Pt foil.

Energies (referenced to the Fermi level) of the K-32, $L_{1-3}$-9.4, and $M_{1-3}$-9.4 conversion electron lines from the deexcitation of $^{83m}$Kr (generated in the EC decay chain of $^{83}$Sr, see Fig.1) measured with the $^{83}$Sr source prepared by ion implantation into the Pt foil are shown in Table 1. In Table 2, the derived (with the use of these conversion line energies) electron binding energies in Kr incorporated into the Pt foil are compared with the experimental values for gaseous krypton (referenced to the vacuum level). Electron binding energies measured for the $M_3$-subshell of Kr implanted in Cu and Ag metals (previous works), and Pt (this work) are displayed in Table 3. In Tables 4 and 5, electron binding energies for some subshells of Kr in different environments related to those of the $L_1$ and K ones, respectively, are given. Experimental natural widths of the K, $L_{1,2,3}$, and $M_1$ atomic levels of krypton in different surroundings are compared in Table 6. Measured and calculated (also in this work) energies and relative intensities of the KLL

Auger transitions in Kr located in different environments are displayed in Table 7. In Table 8, the energy difference of the $KL_2L_3(^1D_2)$ transitions in Rb and Kr measured in the present work is compared with theoretical predictions.

### 3.1. The krypton K-, L-, and M-subshell electron binding energies

From the electron spectra (see Figs. 3-5) measured with the $^{83}$Sr source prepared by ion implantation into the Pt foil, we determined energies $E_F(i)$ (referenced to the Fermi level) of the K and L, M conversion electron lines of the 32 keV E3 and 9.4 keV M1+E2 nuclear transitions in $^{83m}$Kr ($T_{1/2}$=1.83 h), respectively (see Fig.1). They are presented in Table 1. Using the latest energy values of the above nuclear transitions in $^{83m}$Kr, namely $E_{\gamma,9.4}$ = 9 405.8(3) eV [15] and $E_{\gamma,32}$ = 32 151.7(5) eV [16], the electron binding energies $E_{b,F}(i)$ (referenced to the Fermi level) in krypton in the matrix of the platinum host were obtained using the following equation:

$$E_{b,F}(i) = E_{\gamma,j} + E_{\gamma,j,rec} - E_F(i) - E_{rec}(i) \quad (1)$$

Here, $i$ is the krypton atomic shell/subshell index, $j$ = 9.4 or 32.1 keV, $E_{\gamma,j,rec}$ is the energy of the recoil atom after $\gamma$-ray emission ($E_{\gamma,9.4,rec}$ = 0.002 eV, $E_{\gamma,32.1,rec}$ = 0.007 eV) and $E_{rec}(i)$ is the energy of the recoil atom after conversion electron emission ($E_{rec}(K)$ = 0.12 eV, $E_{rec}(L_1)$ = 0.05 eV). They are compared in Table 2 with the experimental values for gaseous krypton of the compilation [12] and results of the latest evaluation [17] for the K shell and L subshells of gas-phase krypton. These values are, however, referenced to the vacuum level which differs from the Fermi level by the work function of our spectrometer. It is seen from the table that (as expected) the electron binding energies in implanted krypton are considerably lower in magnitude (also due to the different reference level) and, generally, the observed difference increases with increasing atomic shell/subshell. In the case of the gas-phase values [12], the M-subshell differences are higher than the L-subshell ones by about 20% and for the data [17], the $L_{2,3}$ difference exceed the $L_1$ difference by a factor of 1.5. This finding contradicts the conclusion made from results of the above mentioned measurements [1,2] based on the X-ray photoelectron spectroscopy that the binding energy shifts are "essentially identical" for the core levels of the investigated rare gases Ne, Ar, Kr, and Xe implanted at 1 keV in the noble metals Cu, Ag, and Au (the determined average ranges of the implanted atoms at this energy were comparable to the mean escape depth for the investigated photoelectrons and span from 1 to 2 nm). Moreover, a drastic decrease by 6.4(15) eV (compared with 0.8 and 0.9 eV determined in Ref. [2] for Ne and Ar, respectively) of the $M_3$ binding energy in Kr is seen from Table 3 between the Ag [2] and Pt (this work) host matrices (no data are given in Ref. [2] for Au host due to troubles with the Kr implantation). We have no explanation for this observation.

Because of the different reference levels used and their approximations, in Table 4 we compared the relative electron binding energies in Kr for different environments, namely the high purity Pt polycrystalline host matrix of the present work, the evaporated layer of the parent $^{83}$Rb on a naturally oxidized aluminum foil of Ref. [18], and $^{83m}$Kr frozen at 4.2 K onto a highly oriented pyrolytic graphite in Ref. [19]. Values for gaseous krypton of both the compilation of Ref. [12] and the latest evaluation [17] are provided in Table 4 too. Our values were obtained from relative positions of the L- and M-9.4 keV conversion electron lines on the scale of the electron retarding voltage applied [3]. Thus a higher precision was reached. The $L_{1-3}$ and $M_{1-3}$ electron binding energies were related to the $L_1$-subshell. It is seen from the table that experimental values for the L-subshells in different surroundings agree very well with each other and also with the compiled data [12] for gas-phase Kr while the evaluation [17] gives differences lower in magnitude by almost the same value of 4.5 eV (i.e. as if the $L_{2,3}$ levels are located closer to the $L_1$ level in Ref. [17]). Nevertheless, a very good agreement is seen for the $M_1$ subshell among the experimental differences for the $M_{1-3}$ subshells in solids. The $M_{2,3}$ subshells are, however, moving away from the $L_1$ subshell starting from the frozen Kr to implanted Kr into the Pt foil (but the agreement within 2σ is kept). The $M_{1-3}$ subshells as a whole in gaseous Kr [12] are closer to the $L_1$ level in comparison with our data for implanted Kr (but again within 2σ). Thus it generally seems that the $M_{2,3}$ shells are shifted more and more away from the $L_1$ as one goes from the gas-phase Kr through the frozen and "evaporated" forms to Kr in the platinum host matrix. This can be explained by the changes in initial-state potential at the core produced by the compression of the valence wave functions in going from the atoms to solids.

In Table 5, the L and $M_1$ electron binding energies related to the K shell one determined for Kr incorporated into the Pt host matrix and $^{83m}$Kr frozen onto a backing at 4.2 K [19] are compared with values obtained from data for gaseous krypton of both the compilation [12] and the latest evaluation [17]. An agreement within 1σ is seen for corresponding shells, not only between the different solids, but also between the solids and gas-phase Kr, with the exception of the $L_1$ shell value [17] which is higher by 3.7(8) eV (4.6 σ) than ours and by 3.3(11) eV than the gas-phase difference [12]. As mentioned above, the $L_{2,3}$ subshells are closer to the $L_1$ one in Ref. [17] by 4.5(4) eV and 4.1(10) eV than in the present work and compilation [12], respectively (see Table 4). This finding indicates that the $L_1$ electron binding energy [17] in gaseous Kr may be incorrect. From the above agreement observed in Table 5 for the L and $M_1$ electron binding energies related to the K shell, a conclusion can be made that different surroundings have no influence on the relative K, $L_{1-3}$ and $M_1$ binding energies within the uncertainties quoted.

By the above approach, we also determined the K-, $L_1$-, and $M_1$-electron binding energies in krypton generated in the EC decay of $^{83}$Rb in an evaporated rubidium layer on the same type platinum foil. The measured spectrum of the corresponding conversion electron lines of the 9.4 keV nuclear transition is shown in Fig. 4. The obtained values were lower by 0.7(1) eV than those measured for the implanted Kr.

*3.2. Natural widths of the krypton K-, $L_{1,2,3}$, and $M_1$ atomic levels*

From the well resolved and sufficiently intense conversion electron lines, we also extracted their natural widths by two different ways. First, the computer code [14] used for the spectrum evaluation gives them as fitted parameters. In addition, we performed a dedicated evaluation by the computer code [25] using only a part of the measured conversion lines, namely its high-energy slope and as few points below the peak as possible, usually two (to set the line at its proper position but still eliminate the effect of the line asymmetry). These spectrum line fragments were fitted by the Voigt function. In the fitting, the Gaussian half-widths (the absolute instrumental resolution of the spectrometer) was kept fixed. The determined values of the natural widths of the K, $L_1$, $L_2$, $L_3$, and $M_1$ levels of Kr in the host of the platinum are compared in Table 6 with those obtained for: i) Kr arising from the EC decay of $^{83}$Rb evaporated on a naturally oxidized aluminum [18]; ii) frozen Kr on highly oriented pyrolytic graphite [19], and iii) gas-phase Kr [20] (a compilation of experimental data). As can be seen from the table, our values generally agree well with those of Ref. [18] for Kr arising from a rubidium layer evaporated on a naturally oxidized aluminum. They are, however, substantially lower than the values measured in Ref. [19] for frozen Kr for the K, $L_1$, and $M_1$ levels. A considerable difference between the natural widths measured for the gas-phase krypton and Kr in solids is observed for the $L_1$ subshell. However, "a considerable line broadening compared with widths of corresponding lines observed in spectra of free gaseous atoms" as stated in Ref. [2] for Ne, Ar, Kr, and Xe implanted in Cu, Ag, and Au was not observed in our work. Within the quoted errors, we did not find any natural width difference for the given atomic level between the sources prepared by the $^{83}$Sr implantation in the Pt foil and by the vacuum evaporation of $^{83}$Rb on the Pt foil.

*3.3. The krypton KLL Auger spectrum*

In the measured electron spectra from both the decay chain of $^{83}$Sr implanted into the platinum foil and the EC decay of $^{83}$Rb in the evaporated layer on the same type of platinum foil, the KLL Auger lines of krypton were also observed (see Fig. 2). We concentrated only on the determination of the absolute transition energies as the KLL Auger spectrum of Kr was thoroughly studied in Ref. [23]. From the measured spectra we obtained energies of 10 838.5(9) and 10 839.5(10) eV (referenced to the Fermi level) for the $KL_2L_3(^1D_2)$ transition in Kr implanted into Pt and generated in the evaporated layer, respectively, i.e., higher by 1.0(2) eV in the latter case (see also Table 7). The observed increase of the krypton $KL_2L_3(^1D_2)$ transition energy for the evaporated $^{83}$Rb source agrees very well with the difference of +0.8(2) eV determined in our work [24] for the rubidium $KL_2L_3(^1D_2)$ transition energy between the $^{85}$Sr sources prepared by vacuum evaporation and implantation into the same type of platinum foil.

The measured $KL_2L_3(^1D_2)$ transition energy of 10 839.5(10) eV for Kr arisen in the evaporated layer on the Pt backing is higher by 2.1(14) eV than the energy of 10 837.4(10) eV (related to the Fermi level) measured in Ref. [23] for the $KL_2L_3(^1D_2)$ transition in Kr generated in the EC decay of $^{83}$Rb evaporated on a naturally oxidized aluminum foil. This fact indicates that in the case of evaporated sources, the source backing material also plays an import role as regards the "chemical state" of the radioactive atoms emitting the studied electrons.

The semi-empirical energy of 10 824.9 eV [21] for the $KL_2L_3(^1D_2)$ transition in gaseous krypton is lower by 13.6(9) and 14.6(10) eV than our above experimental values for Kr implanted in platinum and generated in the evaporated layer on the Pt foil, respectively. The KLL transition energies [21] were calculated for recombination of the initial K vacancies created by, e.g., the internal conversion in daughter $^{83}$Kr (our case) and are referenced to the vacuum level while the primary K vacancies in our solid-phase sources are produced by the electron capture decay and the measured transition energies are referenced to the Fermi level. The observed differences between the theoretical [21] and experimental transition energies cannot be compensated by corrections of the semi-empirical value for both the solid state effect (increasing the Auger transition energies by 6.2 and 7.5 eV [21] for rubidium and platinum, respectively) and a change of the vacuum reference level to the Fermi one (decreasing the transition energies [21] roughly by the work function of our spectrometer equal to about 5.7 eV) as these corrections nearly cancel each other as a result.

Having the electron binding energies determined in the present work for Kr implanted in the platinum host (Table 2), we modified the semi-empirical $KL_2L_3(^1D_2)$ transition energy [21] following the recommendation given in Ref. [21] (with the exception of accounting for a solid-state correction term not provided in Ref. [21] for our case). The obtained value of 10 829.9(17) eV is higher by 5.0(17) eV than the original prediction [21] but still lower by 8.6(19) eV than the measured transition energy for implanted Kr (see Table 7). The omitted solid-state correction term increases (as mentioned above) the calculated Auger transition energies, e.g., by 6.2 and 7.5 eV for Rb and Pt solids [21], respectively. Thus its consideration as well as the below "atomic structure effect" (increasing transition energies by about 6.5 eV for Kr) should bring the modified value [21] to a better agreement with the experiment.

We also performed calculations of energies and intensities of the KLL Auger transition in krypton (Z=36) following the creation of initial vacancies by both the electron capture decay of the parent $^{83}$Rb isotope and internal conversion of gamma rays in the daughter $^{83}$Kr. The aim was to obtain also an estimation of a possible contribution from the "atomic structure effect" to the above discrepancy between the semi-empirical [21] and experimental $KL_2L_3(^1D_2)$ transition energy values. This effect was revealed for the first time in X-rays (see, e.g., Ref. [26]). Higher observed energies of the K Auger transitions following the EC decay (see, e.g., Ref. [27] and references therein) are explained as a result of additional screening of the daughter nucleus by a "spectator" electron, as the $10^{-16}$-$10^{-17}$ s lifetime of the $1s$ atomic hole produced in the EC decay is so short that the intermediate state has the outer-electron configuration close to the parent atom. Thus in the calculations, we assumed the atomic configuration of neutral krypton for the IC decay while for the EC decay, a 5s electron was added to a neutral Kr atom. These "*ab initio*" multiconfiguration Dirac-Fock (MCDF) calculations used the actual electronic configurations of both the initial and final states in the KLL Auger transitions, including the vacancies. Both relativistic effects and quantum-electrodynamic (QED) corrections were taken into account. Each of the Auger transition energies were evaluated as the difference of the total energies of the initial and final states and are referenced to the vacuum level. The calculations were performed for free atoms, i.e., without accounting for the solid state effects. A more detailed description of the calculations is given in the appendix of Ref. [24]. Results of the calculations are presented in Table 7. It is seen from the table that the "atomic structure effect" increases the $KL_2L_3(^1D_2)$ transition energy in Kr by 6.5 eV. Our transition energy calculated for the IC processes is, however, higher by 5.4 eV than the semi-empirical value [21] though both values are referenced to the vacuum level and concern gas-phase Kr. No explanation is available for this finding

It should be also noted that, as in the case of rubidium [24], relative energies of the basic krypton spectrum components calculated in the present work are substantially greater than both the measured [23] and semi-empirical [21] values with the exception of the lines lying close to the $KL_2L_3(^1D_2)$ one. In other words, our calculated spectrum occupies an energy interval wider by 13.1(7) and 18.9 eV than the measured [23] and calculated [21] ones, respectively. As stated in Ref. [24], no explanation is also available for this finding at present time.

In regard to the relative transition intensities, our calculated values agree well with both the relativistic calculations [22] in intermediate coupling with the configuration interaction and the experimental data [23] (see Table 7).

### 3.4. The energy difference of the KLL transitions in Kr and Rb

As mentioned above, in the EC decay chain of the $^{83}$Sr isotope (see Fig. 1) the KLL Auger electrons of both rubidium and krypton are also emitted. After about four $^{83}$Sr half-lives from the source preparation, the KLL Auger lines of the two elements were clearly observed and well resolved in our spectra (see Fig. 2) measured with the $^{83}$Sr source prepared by the ion implantation into the Pt foil. From the relative position of the measured KLL spectra on the electron retarding voltage scale [3] we determined with higher accuracy the energy difference of the $KL_2L_3(^1D_2)$ transitions in rubidium and krypton to be 601.0(8) eV. It is seen from Table 8 that this value agrees very well with the one determined from the absolute energies of the $KL_2L_3(^1D_2)$ transitions in krypton and rubidium measured with the same $^{83}$Sr source in the present work (Table 7) and in Ref. [24], respectively.

The transition energy difference of 605.0 eV obtained from the MCDF calculations performed in the present work for Kr and in Ref. [24] for Rb is higher by 4.0(8) eV (i.e. by 5σ) than the measured value (see Table 8). For both elements, the MCDF calculations were accomplished for the gas-phase system and for K-vacancy produced by the EC-decay.

The semi-empirical calculations [21] give the corresponding transition energy difference to be 610.5 eV, i.e. higher by 9.5(8) eV than the experimental value (Table 8). However, in this case the comparison is not correct because the $KL_2L_3(^1D_2)$ transition energy [21] for Kr is referenced to the vacuum level and is valid for gas-phase system while that one [21] for Rb was calculated for a solid and is related to the Fermi level. But (as mentioned above) a correction of the krypton $KL_2L_3(^1D_2)$ transition energy [21] for the referenced level and the solid-state effect has almost no influence on the energy value because the particular contributions nearly cancel each other.

A value of 605.5(17) eV for the $KL_2L_3(^1D_2)$ transition energy difference is obtained from the calculations [21] when the modified krypton $KL_2L_3(^1D_2)$ transition energy is used (see Table 8). In this case, however, the krypton $KL_2L_3(^1D_2)$ transition energy is lowered because no correction was performed for the solid-state effect (increasing the transition energies [21]).Thus its consideration should decrease the above energy difference by several eV but the exact value is unknown for our case (e.g., values of 6.2 and 7.5 eV are given in Ref. [21] for the Rb and Pt solid-state correction terms, respectively). In addition, in our experiment the primary K vacancies in both elements are produced by the EC decay. According to the results of the "*ab initio*" MCDF transition energy calculations performed in the present work and in Ref. [24], the "atomic structure effect" connected with the EC decay increases the KLL transition energies by 6.5 and 7.0 eV for Kr and Rb, respectively. As a result, the "atomic structure effect" increases the $KL_2L_3(^1D_2)$ transition energy difference of the two elements by 0.5 eV. Thus after the both corrections the resulting transition energy difference can well approximates the experimental value. This fact also indicates the meaningfulness of the adaptation of the Auger transition energies calculated for free atoms for the solid state hosts using the corresponding experimental electron binding energies.

## 4. Conclusion

Using the high-resolution-low-energy "nuclear" electron spectroscopy developed in our laboratory, we determined electron binding energies, natural atomic level widths, and the $KL_2L_3(^1D_2)$ transition energies in krypton in two different solid hosts, namely the high purity polycrystalline platinum and the vacuum evaporated rubidium layer on the same type of platinum foil. The results obtained clearly indicate sensitivity of the studied quantities to the environment of the Kr atoms. This fact has substantial importance for, e.g., the development of electron energy calibration standards for the KATRIN neutrino mass experiment [7]. Results of the performed "*ab initio*" multiconfiguration Dirac-Fock calculations of the rubidium KLL transition energies revealed a significant role of the "atomic structure effect" even for krypton.

**Acknowledgement**

The work was partly supported by grants GACR P 203/12/1896, RFFI 13-02-00756, and the Australian Research Council (grant no. DP140103317).

**Table 1**
Energies $E_F(i)$ (in eV) related to the Fermi level of the K and L, M conversion electron lines of the 32.1 and 9.4 keV nuclear transitions in $^{83m}$Kr, respectively, measured with a source prepared by implantation of the $^{83}$Sr ions in Pt foil.

| Conversion line | K-32 | L$_1$-9.4 | L$_2$-9.4 | L$_3$-9.4 | M$_1$-9.4 | M$_2$-9.4 | M$_3$-9.4 |
|---|---|---|---|---|---|---|---|
| $E_F(i)$ | 17 835.3(11)[a] | 7 491.5(8) | 7 685.5(8) | 7 738.2(8) | 9 124.3(8) | 9 196.2(13) | 9 204.6(15) |

[a] Numbers in parentheses represent uncertainties in last digit(s).

**Table 2**
Krypton electron binding energies $E_{b,F}(i)$ (in eV) related to the Fermi level determined from the measured K, L, and M conversion line energies listed in Table 1 and the corresponding nuclear transition energies [15,16] for Kr atoms implanted into the high purity polycrystalline platinum foil. Electron binding energies $E_{b,V}(i)$ for gaseous Kr [12,17] are referenced to the vacuum level.

| Kr shell/subshell (i) | K | L$_1$ | L$_2$ | L$_3$ | M$_1$ | M$_2$ | M$_3$ |
|---|---|---|---|---|---|---|---|
| $E_{b,F}(i)$ (this work) | 14 316.4(12)[a] | 1 914.3(9) | 1 720.3(9) | 1 667.6(9) | 281.5(9) | 209.6(13) | 201.2(15) |
| $E_{b,V}(i)$ (gaseous Kr [12]) | 14 327.2(8) | 1 924.6(8) | 1 730.9(5) | 1 678.4(5) | 292.8(3) | 222.2(2) | 214.4(2) |
| Implanted – Gaseous [12] | - 10.8(14) | - 10.3(12) | - 10.6(10) | - 10.8(10) | - 11.3(10) | - 12.6(13) | - 13.2(15) |
| $E_{b,V}(i)$ (gaseous Kr [17]) | 14 327.26(4) | 1 921.4(3) | 1 731.91(3) | 1 679.21(3) | | | |
| Implanted – Gaseous [17] | - 10.9(12) | - 7.1(9) | - 11.6(9) | - 11.6(9) | | | |

[a] Numbers in parentheses represent uncertainties in last digit(s).

**Table 3**

Electron binding energies $E_{b,F}(M_3)$ (in eV) referenced to the Fermi level as measured for the $M_3$-subshell of Kr implanted in Cu, Ag [2] and Pt (this work).

| Host matrix | Cu [2] | Ag [2] | Pt (this work) |
|---|---|---|---|
| $E_{b,F}(M_3)$ | 207.27(6) [a)] | 207.60(5) | 201.2(15) |

[a)] Numbers in parentheses represent uncertainties in last digit(s).

**Table 4**

The $L_{2,3}$ and $M_{1-3}$ subshell electron binding energies (in eV) of Kr referenced to the $L_1$ subshell for different solids: a high purity Pt host matrix (this work); an evaporated layer of the parent $^{83}$Rb on a naturally oxidized aluminum foil (Ref [18]); $^{83m}$Kr frozen onto a highly oriented pyrolytic graphite at 4.2 K (Ref. [19]). Values for gaseous krypton were taken from the compilation [12] and the latest evaluation [17].

| Subshell | Experiment | | | Compilation/Evaluation | |
|---|---|---|---|---|---|
| | This work | Ref. [18] | Ref. [19] | Ref. [12] | Ref. [17] |
| $L_1$ | 0.0 | 0.0 | 0.0 | 0.0 | 0.0 |
| $L_2$ | 194.0(2) [a)] | 193.8(2) | 193.7(6) | 193.7(9) | 189.5(3) |
| $L_3$ | 246.6(2) | 246.4(2) | 246.4(6) | 246.2(9) | 242.2(3) |
| $M_1$ | 1 632.8(2) | 1 633.1(8) | 1 632.8(6) | 1 631.8(9) | --- |
| $M_2$ | 1 704.7(8) | 1 703.5(8) | 1 702.7(6) | 1 702.4(8) | --- |
| $M_3$ | 1 713.2(11) | 1 711.1(9) | 1 710.7(6) | 1 710.2(8) | --- |

[a)] Numbers in parentheses represent uncertainties in last digit(s).

**Table 5**

The L and $M_1$ electron binding energies (in eV) referenced to the K shell one measured for Kr in the high purity Pt host matrix (this work) and $^{83m}$Kr frozen onto a carbon backing at 4.2 K [19]. Data for gaseous Kr were obtained from the compilation [12] and the latest evaluation [17].

|  | Experiment | | Gaseous Kr | |
| --- | --- | --- | --- | --- |
| Shell | This work | Ref. [19] | Ref. [12] | Ref. [17] |
| K | 0.0 | 0.0 | 0.0 | 0.0 |
| $L_1$ | 12 402.2(7) [a] | 12 401.3(9) | 12 402.6(11) | 12 405.9(3) |
| $L_2$ | 12 595.9(7) | 12 595.0(9) | 12 596.3(9) | 12 595.4(3) |
| $L_3$ | 12 648.8(7) | 12 647.7(9) | 12 648.8(9) | 12 648.1(1) |
| $M_1$ | 14 035.0(7) | 14 034.1(9) | 14 034.4(9) | ------ |

[a] Numbers in parentheses represent uncertainties in last digit(s).

**Table 6**

Natural widths (in eV) of some atomic levels of krypton in different surroundings: this work – high purity Pt foil; Ref. [18] - naturally oxidized aluminum foil; Ref. [19] - gaseous Kr frozen at 4.2 K on a highly oriented pyrolytic graphite; Ref. [20] (a compilation) – gaseous Kr.

| Atomic level | K | $L_1$ | $L_2$ | $L_3$ | $M_1$ |
| --- | --- | --- | --- | --- | --- |
| This work | 2.3(2) [a] | 4.6(2) | 1.7(2) | 1.3(2) | 3.2(3) |
| Ref. [18] | 2.5(2)/2.4(1) [c] | 4.5(3)/4.5(1) [c] | 1.0(4)/0.8(2) [c] | 1.2(2)/1.2(1) [c] | 2.7(2)/3.0(3) [c] |
| Ref. [19] | 2.83(12) | 5.30(4) | 1.84(5) | 1.40(2) | 4.27(5) |
| Ref. [20] [b] | 2.7(3) | 3.8(9) | 1.3(4) | 1.2(4) | 3.5(4) |

[a] Numbers in parentheses represent uncertainties in last digit(s).
[b] Interpolated values from experimental data.
[c] A value obtained from our re-evaluation of the conversion electron spectra measured in Ref. [18].

**Table 7**

Measured and calculated energies (in eV) and relative intensities ($KL_iL_j/\Sigma KLL$, %) of the KLL Auger transitions in Kr located in different environments: gaseous Kr [21,22,24]; Kr generated in the EC decay of the parent [83]Rb evaporated on a naturally oxidized aluminum foil [23]. EC - the initial vacancies produced by the electron capture decay; IC - the initial vacancies produced by internal conversion.

| Transition | Energy (eV) | | | | Intensities (%) | | |
| --- | --- | --- | --- | --- | --- | --- | --- |
| | Experiment | | Theory | | Experiment | Theory | |
| | Ref. [23] | This work | Ref. [21] | This work [24] | Ref. [23] | This work [24] | Ref. [22] |
| $KL_1L_1(^1S_0)$ | - 430.4(6)[a] | | - 426.7 | - 443.7 | 7.6(3) | 8.0 | 7.7 |
| $KL_1L_2(^1P_1)$ | - 241.7(4) | | - 240.0 | - 248.9 | 15.2(3) | 16.1 | 16.1 |
| $KL_1L_2(^3P_0)$ | - 214.1(7) | | - 208.5 | - 217.8 | 1.7(2) | 1.7 | 1.7 |
| $KL_1L_3(^3P_1)$ | - 181.6(4) | | - 178.7 | - 186.6 | 6.8(2) | 6.6 | 6.3 |
| $KL_1L_3(^3P_2)$ | - 159.6(5) | | - 156.2 | - 163.5 | 3.4(2) | 3.6 | 3.4 |
| $KL_2L_2(^1S_0)$ | - 58.2(5) | | - 55.4 | - 58.4 | 3.5(2) | 3.5 | 3.5 |
| $KL_2L_3(^1D_2)$ | 10 837.4(10)[b] EC | 10 838.5(9)[b,c] EC<br>10 839.5(10)[b,d] EC | 10 824.9[e] IC<br>10 829.9(17)[b,f] IC | 10836.8[e] EC<br>10830.3[e] IC | 47.4(8) | 46.3 | 47.0 |
| $KL_3L_3(^3P_0)$ | + 42.7(4) | | + 42.6 | + 43.3 | 2.9(2) | 2.8 | 2.8 |
| $KL_3L_3(^3P_2)$ | + 62.5(3) | | + 60.4 | + 62.3 | 11.5(2) | 11.4 | 11.5 |

[a] Numbers in parentheses represent uncertainties in last digit(s).
[b] Referenced to the Fermi level.
[c] Kr implanted into the high purity Pt foil.
[d] Kr generated in the EC decay of the parent [83]Rb evaporated on the high purity Pt foil.
[e] Referenced to the vacuum level (gas-phase Kr).
[f] A value obtained in the present work by the use of the relevant electron binding energies from Table 2 for Kr in the platinum host.

**Table 8**
A comparison of the energy differences (in eV) of the $KL_2L_3(^1D_2)$ transitions in Rb and Kr directly measured in the present work for the $^{83}$Sr source prepared by ion implantation into the platinum foil with values obtained from results of the semi-empirical calculations [21] (for the IC processes) and the "*ab initio*" MCDF calculations of the present work and Ref. [24] (for the EC decay).

| Experiment | | Theory | | |
|---|---|---|---|---|
| This work | This work [a] | Ref. [21] | Ref. [21] [b] | This work and Ref. [24] |
| 601.0(8) [c] | 601.8(14) | 610.5 | 605.5(17) | 605.0 |

[a] The value obtained from the absolute $KL_2L_3(^1D_2)$ transition energies of Kr (this work) and Rb (Ref. [24]) measured for the same $^{83}$Sr source.

[b] The krypton $KL_2L_3(^1D_2)$ transition energy [21] modified by the use of the electron binding energies determined in the present work (Table 2) without accounting for the solid state effect.

[c] Numbers in parentheses represent uncertainties in last digit(s).

**Figure captions**

Fig. 1 The incomplete decay schemes of the $^{83}$Sr and $^{83}$Rb isotopes [4].

Fig. 2 An overview low-energy electron spectrum emitted in the $^{83}$Sr decay measured with 21 eV instrumental resolution and 7 eV step and with 30 s exposition time per spectrum point after four $^{83}$Sr half-lives from the source preparation. The spectrum was not corrected for the $^{83}$Sr decay and for the spectrometer transmission drop [3,8] with increasing electron retarding voltage. In the spectrum, electrons following the EC decay of $^{83}$Sr to $^{83}$Rb and $^{83}$Rb to $^{83}$Kr are seen. In the insert, a spectrum region including a part of the KLL Auger spectrum of Kr and the full KLL Auger spectrum of Rb is shown on an enlarged scale.

Fig. 3 The L and M subshell conversion electron lines of the 9.4 keV M1+E2 nuclear transition in $^{83}$Kr emitted in the EC decay chain of of $^{83}$Sr (see Fig. 1). The source used was prepared by $^{83}$Sr ion implantation into a high purity polycrystalline Pt foil at 30 keV. The spectrum was measured after twelve $^{83}$Sr half-lives from the source preparation with 7 eV instrumental resolution and 2 eV step size in 8 sweeps. The exposition time per spectrum point was 60 s. A decomposition of the measured spectrum into individuals are displayed by continuous lines. The "discrete energy loss" peaks (see section 2.3. for explanation) are denoted by DEL. The $M_{1-3}$-9.4 conversion lines are shown in the insert in the enlarge scale.

Fig. 4 The L and M subshell conversion electron lines of the 9.4 keV M1+E2 nuclear transition in $^{83}$Kr emitted in the EC decay of $^{83}$Rb. The $^{83}$Rb source was prepared by vacuum evaporation on a high purity polycrystalline Pt foil. The spectrum was measured with instrumental resolution of 7 eV and 2 eV step in 4 sweeps. The exposition per spectrum point in each sweep was 60 s. Results of spectrum decomposition are shown by continuous lines. DEL means "discrete energy loss" peak (see section 2.3.). The insert shows, on an enlarged scale, the $M_{1-3}$-9.4 conversion line group.

Fig. 5 Examples of the K conversion electron line of the 32 keV E3 nuclear transition in $^{83m}$Kr (generated in the EC decay of $^{83}$Rb, see Fig. 1) taken at instrumental resolution of 7 eV with 2 eV step size. The $^{83}$Rb sources used were prepared by vacuum evaporation on the Pt foil (full circles) and by implantation of $^{83}$Sr ions at 30 keV into the same foil (open circles). The "discrete energy loss" peak (see section 2.3.) is denoted by DEL (upper spectrum).

Figure 1

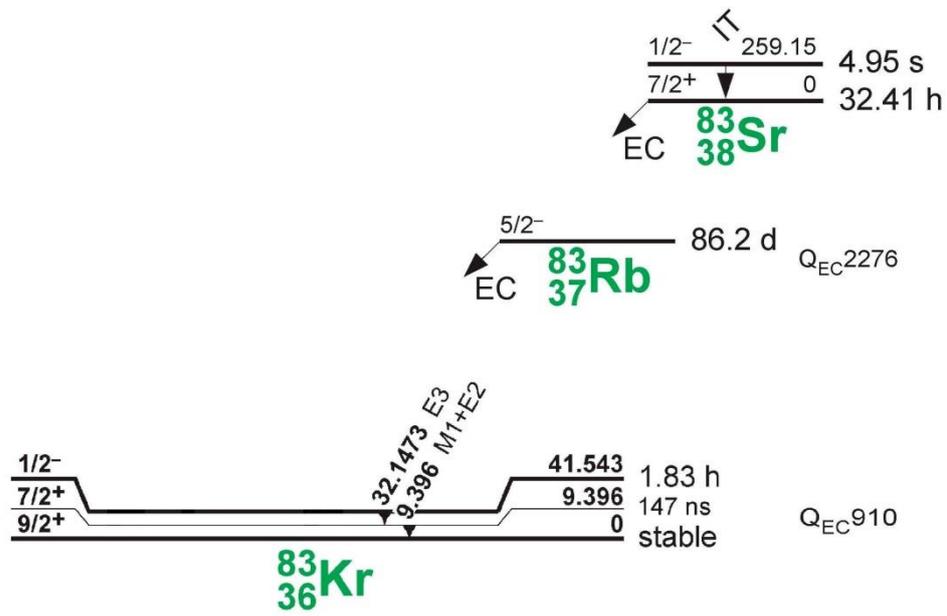

Figure 2

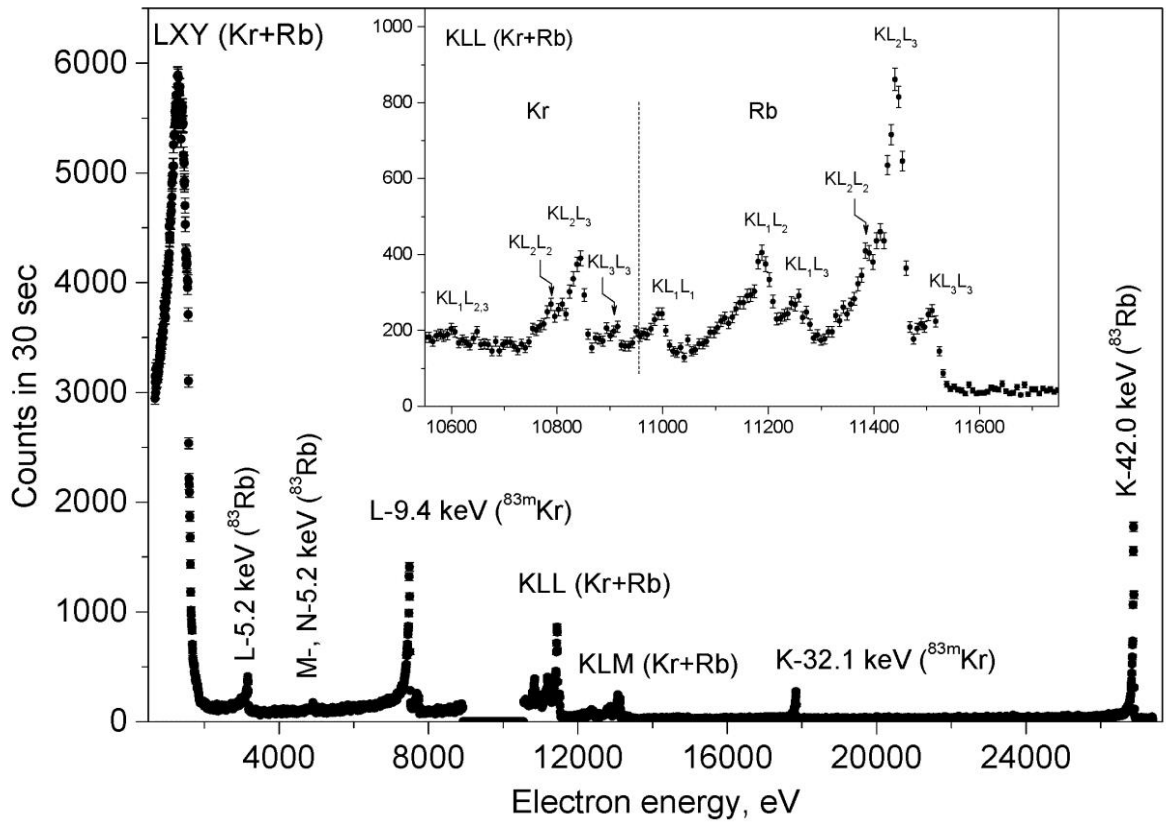

Figure 3

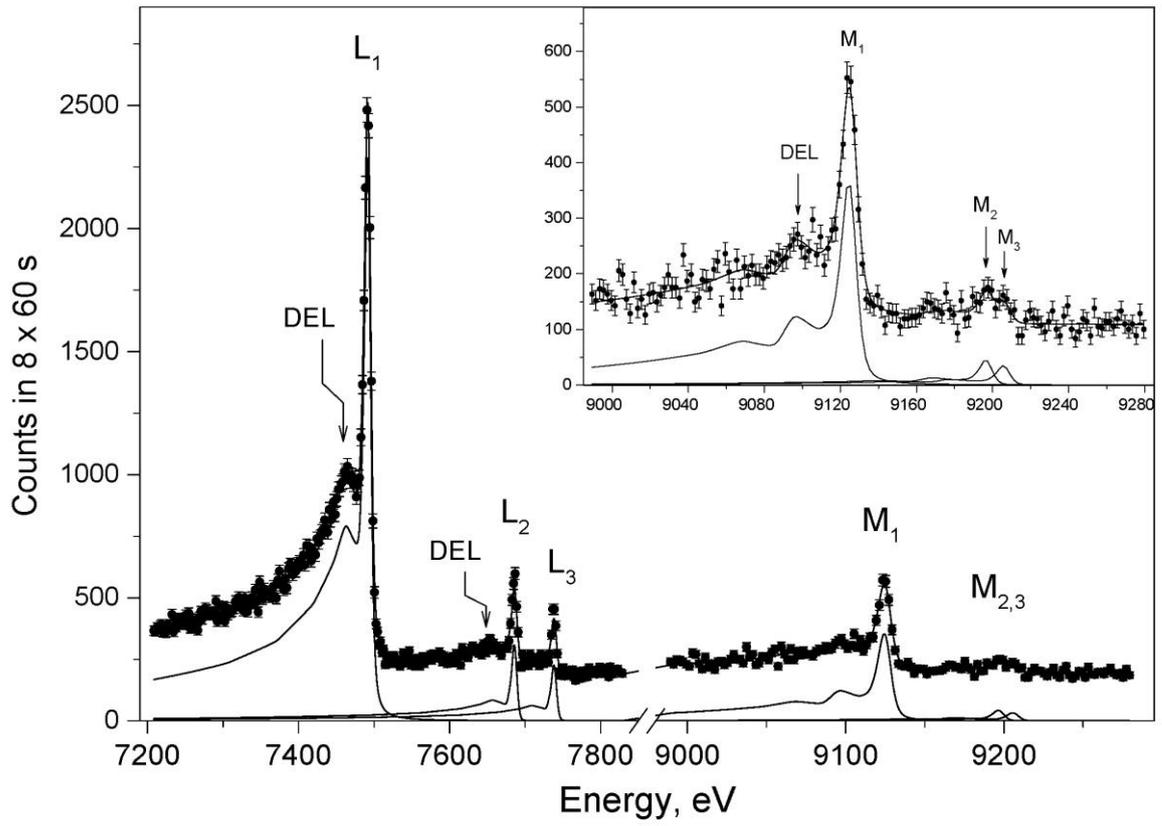

Figure 4

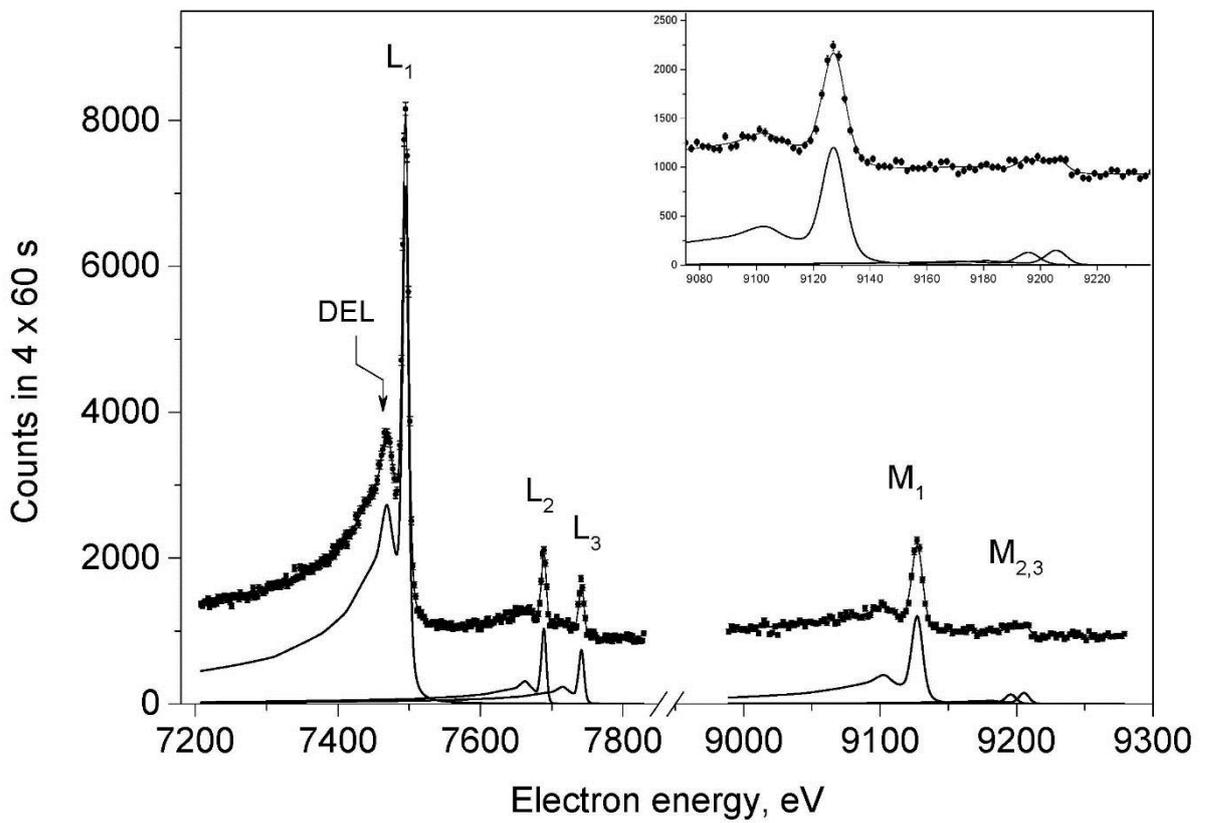

Figure 5

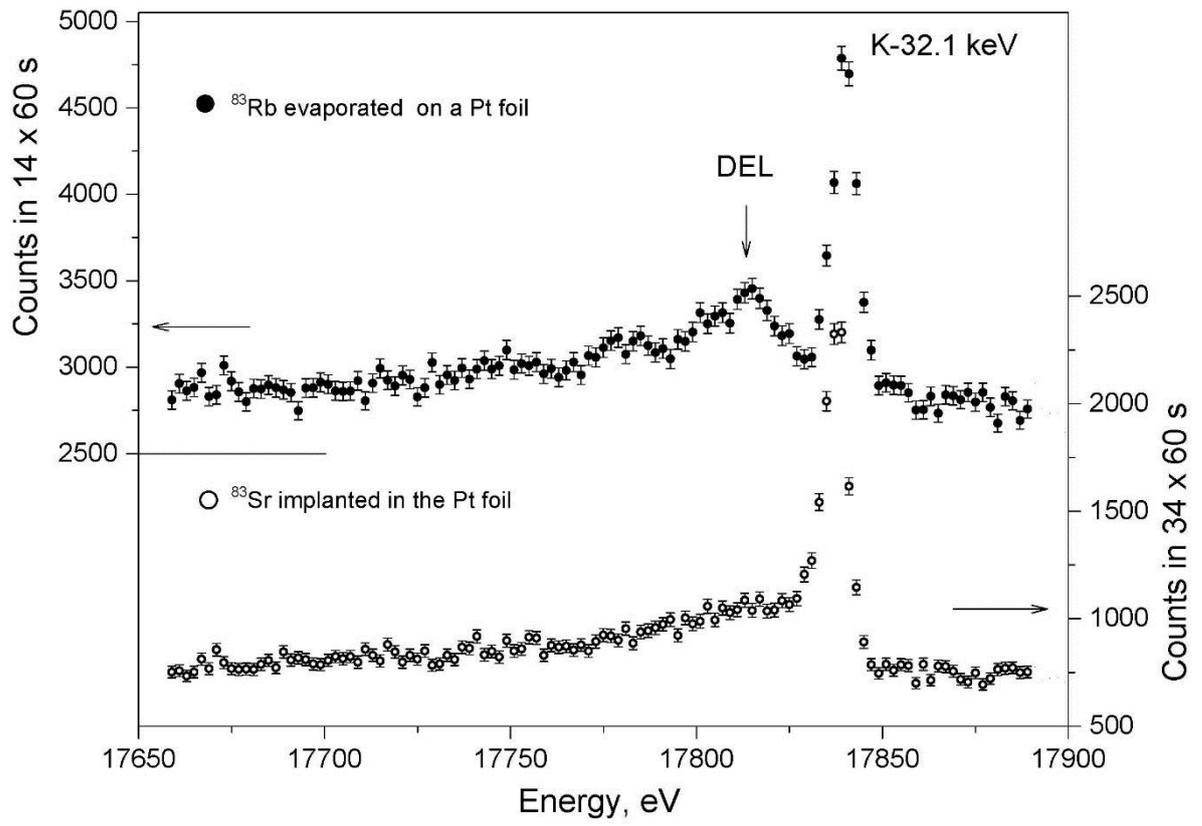